\documentclass[aps,prl,preprint]{revtex4}
\usepackage{epsfig}

\begin{document}

\title{MAGNETIZATION DENSITIES AS REPLICA PARAMETERS:
THE DILUTE FERROMAGNET}

\author{Maurizio Serva}  
\affiliation{Dipartimento di Matematica,
Universit\`a dell'Aquila,
I-67010 L'Aquila, Italy}

\bigskip

\date{\today}

\begin{abstract}

In this paper we compute exactly the
ground state energy and entropy of
the dilute ferromagnetic Ising model.
The two thermodynamic quantities are also computed 
when a magnetic field with random locations is present.
The result is reached in the replica approach
frame by a class of replica order parameters
introduced by Monasson~\cite{M}.
The strategy is first illustrated considering the SK model,
for which we will show the complete
equivalence with the standard replica approach.
Then, we apply to the diluted ferromagnetic Ising model
with a random located magnetic field,
which is mapped into a Potts model.
This formalism can be, in principle,
applied to all random systems, and we believe
that it could be of help in many other contexts.

\bigskip
\noindent
PACS number͑s͒: 05.50.+q, 64.60.De, 75.10.Hk, 81.05.Kf

\noindent
Keywords: disordered systems, replica trick, dilute ferromagnet
\end{abstract}

\maketitle

\section{1 - Introduction}

The theoretical modeling of statistical systems in many areas of 
physics makes use of random Hamiltonians. 
Assuming self-averaging, observable quantities may be evaluated by 
the logarithmic average of the partition function $<$$\log(Z)$$>$ over the
quenched random variables.  
In this way, the technically difficult
task of computing $Z$ 
for a given realization of the disorder is avoided.  
However, the mathematical operation of directly computing 
$<$$\log(Z)$$>$ is also difficult, 
and can be done only using replica trick, 
which implies the computation of  $<$$Z^n$$>$.

Despite its highly successful application to the treatment
of some disordered systems 
(the most celebrated is the SK model~\cite{SK,MPV}, 
the replica trick encounters serious difficulties when applied 
to many other models.
The main reason is the unbounded proliferation of replica order 
parameters, as for example the multi-overlaps 
in dilute spin glass~\cite{FL,GT}. 

In this paper we use a class of replica order 
parameters firstly introduced by Monasson~\cite{M},
which, in principle can be used for all models.
We preliminarily observe that the computation of $<$$Z^n$$>$
implies the sum over all the realizations of the spin variables 
$\sigma_i = (\sigma_i^1,\sigma_i^2,......,\sigma_i^n )$,
any of them may take $2^n$ values.
Assume that $x(\sigma)N$ is the number of vectors $\sigma_i$ 
which equal the given vector $\sigma$ =
(${\sigma^1,\sigma^2,........,\sigma^n }$),
then, $<$$Z^n$$>$ can be re-expressed in terms of a sum
over the possible positive values of the $2^n$ 
magnetization densities
$x(\sigma)$, with the constraint $\sum_{\sigma} x(\sigma)=1$.
In practice, this is equivalent to maximize  
with respect to these order parameters.

The strategy proposed here will be illustrated, 
considering the SK model, in next section, where we will also 
show the complete equivalence with the standard approach.
Nevertheless, all models can be described in terms of the 
$2^n$ order parameters $x(\sigma)$.
We will tackle the dilute ferromagnetic Ising model
in section 3 and in section 4 we will compute exactly the
ground state energy and entropy.
The results coincide with those found in \cite{DSG}
where the approach is not based on replicas.
In section 5 we extend the scope by considering the same model
in presence of a magnetic field with random locations,
and, finally, in section 6 we are able to compute 
the ground state energy and entropy also in this case.
Conclusions and outlook are in the final section.

\section{2 - SK spin glass}

In this section we try to illustrate our approach 
considering the SK model. 
We do not have new results concerning SK, 
but we just show the complete equivalence with
the standard replica approach. 

The partition function is

\begin{equation}
Z= \sum_{\#} \exp \left( \frac{\beta}{\sqrt{N}}\sum_{i>j}
J_{ij} \sigma_i \sigma_j \right)
\label{skzeta}
\end{equation}
where the sum $\sum_{\#}$ goes over the $2^{N}$ realizations of the
 ${\sigma_i}$, the sum $\sum_{i>j}$ goes over all $N(N-1)/2$ 
pairs $\, i  j$ and the $J_{ij}$ are independent random 
variables with 0 mean and variance 1.
Then, by replica approach, neglecting terms which vanish in the
thermodynamic limit, we have

\begin{equation} 
<Z^n> = \sum_{\#} \exp \left( \frac{\beta^2}{4N} \sum_{i,j} \,
( \sum_{\alpha} \sigma_i^\alpha \sigma_j^\alpha)^2 \,\right)
\label{sknzeta}
\end{equation}
where the sum $\sum_{\#}$ goes over the $2^{nN}$ realizations of 
the ${\sigma_i^\alpha}$ and the sum $\sum_{i,j}$ goes over all $N^2$
possible values of $i$ and $j$.

Assume that $Nx(\sigma)$ is the number of vectors
(${\sigma_i^1,\sigma_i^2,......,\sigma_i^n }$) 
which equal the given vector $\sigma$ =
(${\sigma^1,\sigma^2,.......,\sigma^n }$),  
then, according to~\cite{M}, we can write

\begin{equation} 
\sum_{i,j} \, \left( \sum_{\alpha} \sigma_i^\alpha \sigma_j^\alpha \right)^2 \,
=N^2 \, \sum_{\sigma, \tau} x(\sigma) x(\tau) (\sigma \tau )^2 
\label{sken}
\end{equation}
where $\sum_{\sigma, \tau}$ goes over the $2^{2n}$
possible values of the variables $\sigma$ and $\tau$
and where $\sigma \tau$ is the scalar product
$\sigma \tau = \sum_{\alpha} \sigma^{\alpha} \tau^{\alpha}$.
Then, observe that the number of realizations corresponding 
to a given value of the $2^n$ magnetization
densities $x(\sigma)$ is 

\begin{equation} 
\exp(-N \,\sum_{\sigma} x(\sigma)\log(x(\sigma))
\label{skentr}
\end{equation}
where $\sum_{\sigma}$ is the sum over the $2^n$
possible values of the variable $\sigma$.
Indeed, in the above expression we neglected
terms which are in-influent in the thermodynamic limit.

We can now define $\Phi_n$ as the large $N$ limit of 
$\frac{1}{N} \log<$$Z^n$$>$, then
 
\begin{equation} 
\Phi_n = \max_{x}  \;  [ \ \frac{\beta^2}{4} 
\sum_{\sigma, \tau} x(\sigma) x(\tau) (\sigma \tau )^2 
-\sum_{\sigma} x(\sigma) \log( x(\sigma)) \, ]
\label{skphin}
\end{equation}
The maximum is taken over the possible
values of the $2^n$ order parameters $x(\sigma)$
provided that $\sum_{\sigma} x(\sigma) =1$.
The constraint can be accounted by adding the Lagrangian multiplier
$\lambda ( \sum_{\sigma}x(\sigma)-1)$ to expression (\ref{skphin}),
then the maximum is given by

\begin{equation} 
 \frac{\beta^2}{2} \sum_{\tau} x(\tau) (\sigma \tau )^2 
 -\log( x(\sigma)) -1= \lambda
\label{skl}
\end{equation}
where $\lambda$ has to be chosen in order to
have  $\sum_{\sigma} x(\sigma) =1$.
From this equation we get that the maximum is realized for the set
of the $2^n$ order parameters $x(\sigma)$ which satisfy

\begin{equation} 
 x(\sigma) =   \frac{1}{A}
 \exp\left( \frac{\beta^2}{2}
 \sum_{\tau} x(\tau) (\sigma \tau )^2 \right)
\label{skx}
\end{equation}
where the sum $\sum_{\tau}$ goes over the 
$2^n$ possible values of the variable $\tau$ and where $A$ is

\begin{equation} 
A =\sum_{\sigma}
 \exp\left( \frac{\beta^2}{2}
 \sum_{\tau} x(\tau) (\sigma \tau )^2 \right)
\label{skA}
\end{equation}
and where $\sum_{\sigma}$ is the sum over the $2^n$
possible values of the variable $\sigma$.

The explicit solution could be found, in principle, 
by a proper choice of the parametrization of
the $x(\sigma)$, but it can be easily seen that this solution
coincides with the standard solution
of the SK model.
In fact, if one defines the symmetric matrix $q_{\alpha\beta}$ as

\begin{equation} 
 q_{\alpha \beta} =
 \sum_{\tau} x(\tau) \tau^\alpha \tau^\beta
\label{skq}
\end{equation}
one gets

\begin{equation}  
x(\sigma) = \frac{1}{\tilde{A}}
\exp \left( \beta^2\ \sum_{\alpha>\beta} 
q_{\alpha \beta}\sigma^{\alpha}\sigma^{\beta} \right)
\label{skqx}
\end{equation}
where

\begin{equation}  
\tilde{A} = 
\sum_{\sigma} \exp\left( \beta^2 \ \sum_{\alpha>\beta} 
q_{\alpha \beta}\sigma^{\alpha}\sigma^{\beta}\right)
\label{skqA}
\end{equation}
the sum $\sum_{\alpha>\beta} $ goes over all $n(n-1)/2$ pairs
$\alpha, \beta$ and 
the diagonal therms of the matrix $q_{\alpha\beta}$
disappeared since they cancel out
in the expressions (\ref{skx}), (\ref{skA}).

Then, after some work, the expression (\ref{skphin}) rewrites as

\begin{equation} 
\Phi_n =  \; \max_{q}  \;  [ \,
\frac{\beta^2}{4} \, n  
- \frac{\beta^2}{2}
\sum_{\alpha>\beta} \left( q_{\alpha \beta}  \right)^2 
+ \log \left(\sum_{\sigma} \exp( \beta^2 \ \sum_{\alpha>\beta} 
q_{\alpha \beta}\sigma^{\alpha}\sigma^{\beta})\right) \, ]
\label{skstandard}
\end{equation}
where the maximum is over the variables $q_{\alpha \beta}$.
This is the standard solution of~\cite{SK} provided
the proper maximum of $q_{\alpha \beta}$ is found~\cite{MPV}.
Then $\Phi =  \lim_{n \to 0}\frac{\Phi_n}{n} = S-\beta E $
where $S$ is the entropy and $E$ the energy.

\section{3 - Dilute ferromagnet}

In this section, we show how our approach works   
for the dilute ferromagnetic system.
This model is much less studied than SK, and many informations
about its phenomenology are still missing.

The partition function is

\begin{equation}
Z= \sum_{\#} \exp \left( \beta \sum_{i>j} K_{ij} \sigma_i \sigma_j \right)
\label{fz}
\end{equation}
where $K_{ij}$ are quenched variables
which take the value $1$ with probability $\frac{\gamma}{N}$
and $0$ otherwise.
The dilution coefficient $\gamma$ may take any positive value
and the number of bonds is about $\frac{\gamma N}{2}$.
This model has been recently studied in a recent paper~\cite{GT},
while models with lesser dilution~ \cite{BC,BG} ( number of bonds of
order $ N^\epsilon $ with $\epsilon >1$ ) have been also considered.

We can rewrite the above expression as

\begin{equation}
Z= \exp \left( \frac{\gamma N}{2} \log(\cosh(\beta))\right)
\sum_{\#} \prod_{i>j} \left(1+ \tanh(\beta)
K_{ij}\sigma_i \sigma_j \right)
\label{fz2}
\end{equation}
where the equality holds in the sense that
$\frac{1}{N}\log(Z)$ coincide in the thermodynamic limit
for (\ref{fz}) and (\ref{fz2}) because $\sum_{i>j} K_{ij}
= \frac{\gamma}{2}N + o(N)$

Let us define
\begin{equation}
G= \sum_{\#} \prod_{i>j} \left(1+ \tanh(\beta)
K_{ij}\sigma_i \sigma_j \right)
\label{fg}
\end{equation}
than we can compute $<$$G^n$$>$

\begin{equation}
<G^n> = \sum_{\#} \prod_{i>j} 
\left(1- \frac{\gamma}{N} + \frac{\gamma}{N}
\prod_\alpha(1+\tanh(\beta) 
\sigma_i^\alpha \sigma_j^\alpha)\right)
\label{fgn}
\end{equation}
which is thermodynamically equivalent to

\begin{equation}
<G^n> = \exp ( -\frac{\gamma N}{2})
\sum_{\#} 
\exp  \frac{\gamma}{N} \sum_{i>j}
\prod_\alpha\left(1+\tanh(\beta) 
\sigma_i^\alpha \sigma_j^\alpha \right)
\label{fgn2}
\end{equation}

By introducing the $x(\sigma)$
with identical meaning as in previous section,
and defining $\Psi_n$ as the large $N$ limit of 
$\frac{1}{N} \log<$$G^n$$>$ we get

\begin{equation}
\Psi_n = \max_{x}  \;  [ \, -\frac{\gamma}{2} + \frac{\gamma}{2}\sum_{\sigma,\tau} 
x(\sigma) x(\tau)
\prod_\alpha  (1+\tanh(\beta) 
\sigma^\alpha \tau^\alpha)
-\sum_{\sigma} x(\sigma) \log( x(\sigma)) \, ]
\label{gnlog}
\end{equation}
where the maximum is taken with respect the $2^n$
order parameter $x(\sigma)$. To obtain an explicit expression 
one should find their parametric expression.
This will be done in next section only for the zero
temperature case. Formally, the maximum is reached for

\begin{equation} 
 x(\sigma) =   \frac{1}{A}
 \exp\left( \gamma
 \sum_{\tau} x(\tau)  
\prod_\alpha  (1+\tanh(\beta) 
\sigma^\alpha \tau^\alpha) \right)
\label{fx}
\end{equation}
where

\begin{equation} 
A =\sum_{\sigma}
\exp\left(\gamma
 \sum_{\tau} x(\tau)  
\prod_\alpha  (1+\tanh(\beta) 
\sigma^\alpha \tau^\alpha) \right)
\label{fA}
\end{equation}
Then, according to  (\ref{fz2}),  (\ref{fg}), (\ref{fgn2}) 
and  (\ref{gnlog}), we get

\begin{equation}
\Phi =  \lim_{n \to 0}\frac{\Psi_n}{n}+
\frac{\gamma}{2}\log(\cosh(\beta)) = S-\beta E 
\label{fpsi}
\end{equation}
where $S$ is the entropy and $E$ the energy.

\section{4 - Dilute ferromagnet: zero temperature}

Let us consider the simpler case of vanishing temperature.
In this limit $\tanh(\beta)=1$
and expression (\ref{gnlog}) becomes

\begin{equation}
\Psi_n= \max_{x}  \;  [ \,
-\frac{\gamma}{2}+ \frac{\gamma}{2} 2^n \sum_{\sigma} 
x(\sigma)^2 -\sum_{\sigma} x(\sigma) \log( x(\sigma)) \, ]
\label{gtpott}
\end{equation}
which is a standard $2^n$-components Potts model.
The solution is known and can be found assuming that $2^n-1$ 
quantities $x(\sigma)$ take the value $\frac{1-\theta}{2^n}$
and one takes the value $\frac{1+(2^n-1)\theta}{2^n}$.
The state with different value can be any of the possible $2^n$,
we assume that is the one with $\sigma^\alpha=1$ for all $\alpha$.
We can write:

\begin{equation}
x(\sigma) = \frac{1-\theta}{2^n} + 
 \frac{\theta\prod_{\alpha=1}^n (1+\sigma^\alpha)}{2^n}
\label{xxpott}
\end{equation}
which obviously satisfy the constraint $\sum_{\sigma} x(\sigma) =1$.
Inserting the above expression in (\ref{gtpott}) we obtain

\begin{equation}
\Psi_n= \max_{\theta}  \;  [ \,
\frac{\gamma}{2} (2^n -1) \theta^2+  B_n \, ]
\label{ft}
\end{equation}
with

\begin{equation}
B_n =  -\left( \frac{1+(2^n -1)\theta}{2^n} \right)
\log(1+(2^n -1)\theta)
-(2^n-1)\frac{1-\theta}{2^n}
\log(1-\theta)+n\log(2)
\label{fB}
\end{equation}

If we expand to the first order in $n$ the above expression, we obtain

\begin{equation}
\Psi_n= \max_{\theta}  \;  [ \,
n\log(2)\frac{\gamma}{2} 
\, \theta^2+
n\log(2)(1-\theta) (1-\log(1-\theta)) \, ]
\label{fgt}
\end{equation}
Then, if we take into account (\ref{fpsi}) and we 
also take into account that for large $\beta$ 
one has
$\log(\cosh(\beta)) = \beta -\log(2)$, we have 
that the energy at 0 temperature equals 

\begin{equation}
E = -\frac{\gamma}{2}
\label{fe}
\end{equation}
while the entropy $S$ is
$ -\frac{\gamma}{2} \log(2) + \lim_{n \to 0}\frac{\Psi_n}{n}$.
Therefore:

\begin{equation} 
S=\log(2)\frac{\gamma}{2} 
( \theta_c^2-1)+
\log(2)(1-\theta_c) (1-\log(1-\theta_c))
\label{fs}
\end{equation}
where $\theta_c$ is given by the equation

\begin{figure}
\vspace{.2in}
\centerline{\psfig{figure=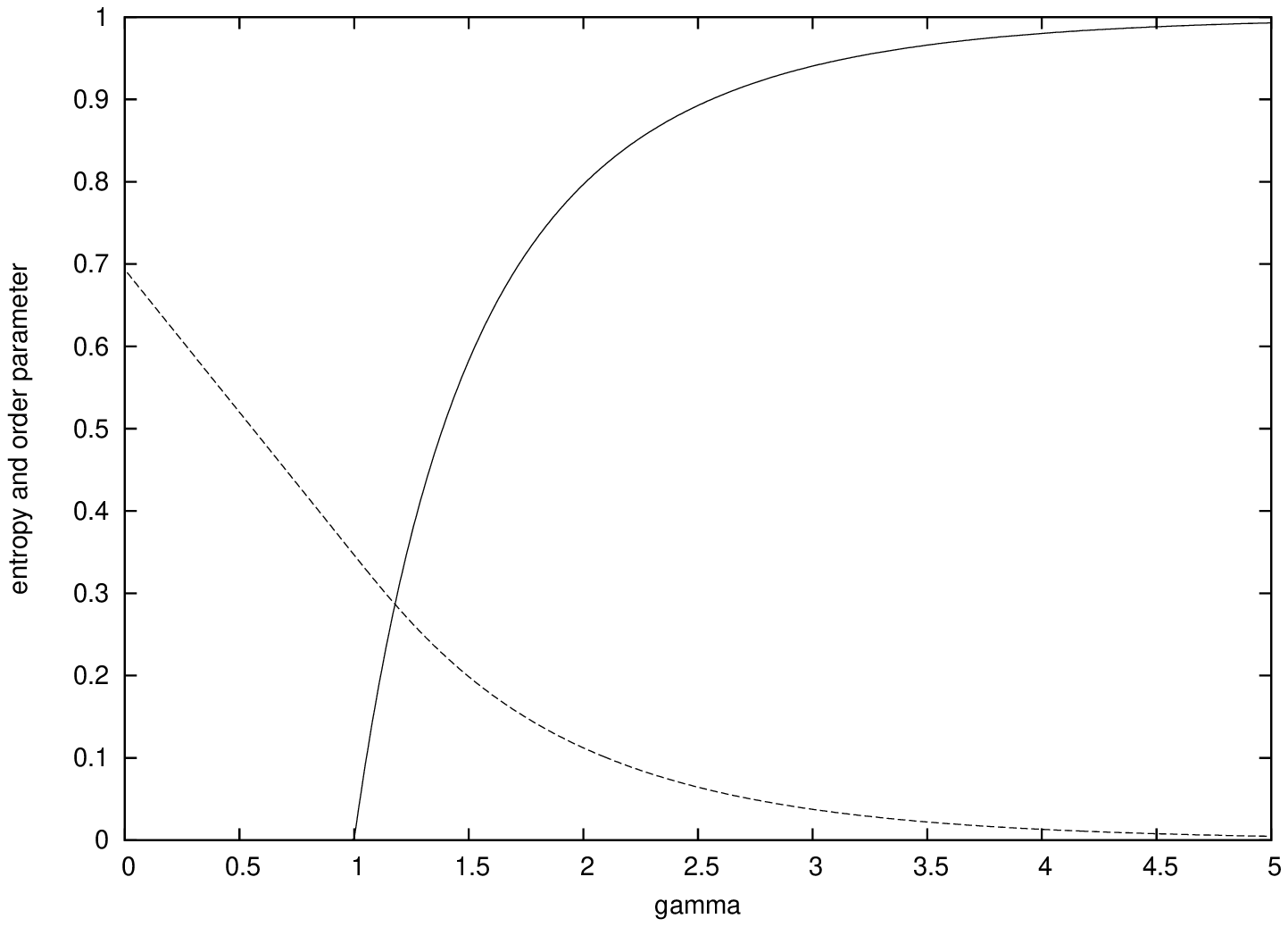,width=4.0truein,angle=270}}
\bigskip
\caption{
Entropy (dashed line) and order parameter $\theta_c$ 
(full line) as a function of
the dilution coefficient $\gamma$ at 0 temperature and 0 magnetic field.
The transition is at $\gamma=1$ where the first derivatives of
both entropy and $\theta_c$ are discontinuous.}
\label{fig1.ps}
\end{figure}

\begin{equation}
\exp(-\gamma\theta_c) = 1-\theta_c
\label{ftheta}
\end{equation}
This equation has a single solution $\theta_c=0 $ 
if  $\gamma \leq 1$
and one more non trivial solution if $\gamma > 1$ which corresponds
to the maximum.
Therefore, at 0 temperature, for  $\gamma < 1$ the system is
in a paramagnetic phase while for $\gamma >1$
is in a disordered ferromagnetic phase.
The transition corresponds to the percolation transition
generated by the ferromagnetic links.
The entropy $S$ and the order parameter
$\theta_c$ are plotted in Fig. 1 as a function of
the dilution coefficient $\gamma$.
At the transition value $\gamma=1$, the first derivatives of
both entropy and $\theta_c$ are discontinuous.

\section{5 - Dilute ferromagnet in a magnetic field}

In this section, we show how our approach works 
for the dilute ferromagnetic system in a magnetic field 
with random locations.

The partition function of this model is

\begin{equation}
Z= \sum_{\#} \exp \left( \beta \sum_{i>j} K_{ij} \sigma_i \sigma_j 
+ \beta \sum_{i} h_{i} \sigma_i  \right)
\label{fzm}
\end{equation}
where $K_{ij}$ are the previously defined quenched variables
and the $h_i$  take the value $h$ with probability $\delta$
and $0$ otherwise.
We can rewrite the above expression as $Z=PG$ where

\begin{equation}
G=  \sum_{\#} \prod_{i>j} \left(1+ \tanh(\beta)
K_{ij}\sigma_i \sigma_j \right)
\prod_{i} \left(1+ \tanh(\beta h_i) \sigma_i \right)
\label{fzm2}
\end{equation}
and where

\begin{equation}
P= \exp \left( \frac{\gamma N}{2} \log(\cosh(\beta))\right)
\exp \left( \delta N  \log(\cosh(\beta h))\right)
\label{pr}
\end{equation}
the equality holds in the sense that
$\frac{1}{N}\log(Z)$ coincide in the thermodynamic limit
when computed from (\ref{fzm}) and 
from $Z=PG$ with $G$ and $P$ given by (\ref{fzm2}) and (\ref{pr}).

If one takes into account that
\begin{equation}
<\prod_{i,\alpha} \left(1+ \tanh(\beta h_{i})
\sigma_i^\alpha \right)> 
= \prod_{i} \left(1-\delta + \delta \prod_\alpha(1+\tanh(\beta h) 
\sigma_i^\alpha)\right)
\label{fgnm}
\end{equation}
one has that $\Psi_n$, with respect to (\ref{gnlog}),
contains the extra term  

\begin{equation}
\sum_{\sigma} x(\sigma) 
\log \left( 1-\delta +\delta \prod_\alpha  (1+\tanh(\beta h)
\sigma^\alpha) \right)
\label{gnlogm1}
\end{equation}
which
is associated to the magnetic field.

Then, if the correct maximum is found
we get
\begin{equation}
\Phi =  \lim_{n \to 0}\frac{\Psi_n}{n}+
\frac{\gamma}{2}\log(\cosh(\beta)) 
+ \delta \log(\cosh(\beta)) = S-\beta E 
\label{fpsim}
\end{equation}

\section{6 - Dilute ferromagnet in a magnetic field: zero temperature}

In the vanishing temperature limit
one has $\tanh(\beta)=\tanh(\beta h)=1$
and expression $\Psi_n$ becomes

\begin{equation}
\Psi_n=  \max_{x}  \;  [
-\frac{\gamma}{2}+ \frac{\gamma}{2} 2^n \sum_{\sigma} 
x(\sigma)^2 + 
\sum_{\sigma} x(\sigma) 
\log( 1-\delta +\delta \prod_\alpha  (1+\sigma^\alpha))
-\sum_{\sigma} x(\sigma) \log( x(\sigma)) \, ]
\label{gtpottm}
\end{equation}

\begin{figure}
\vspace{.2in}
\centerline{\psfig{figure=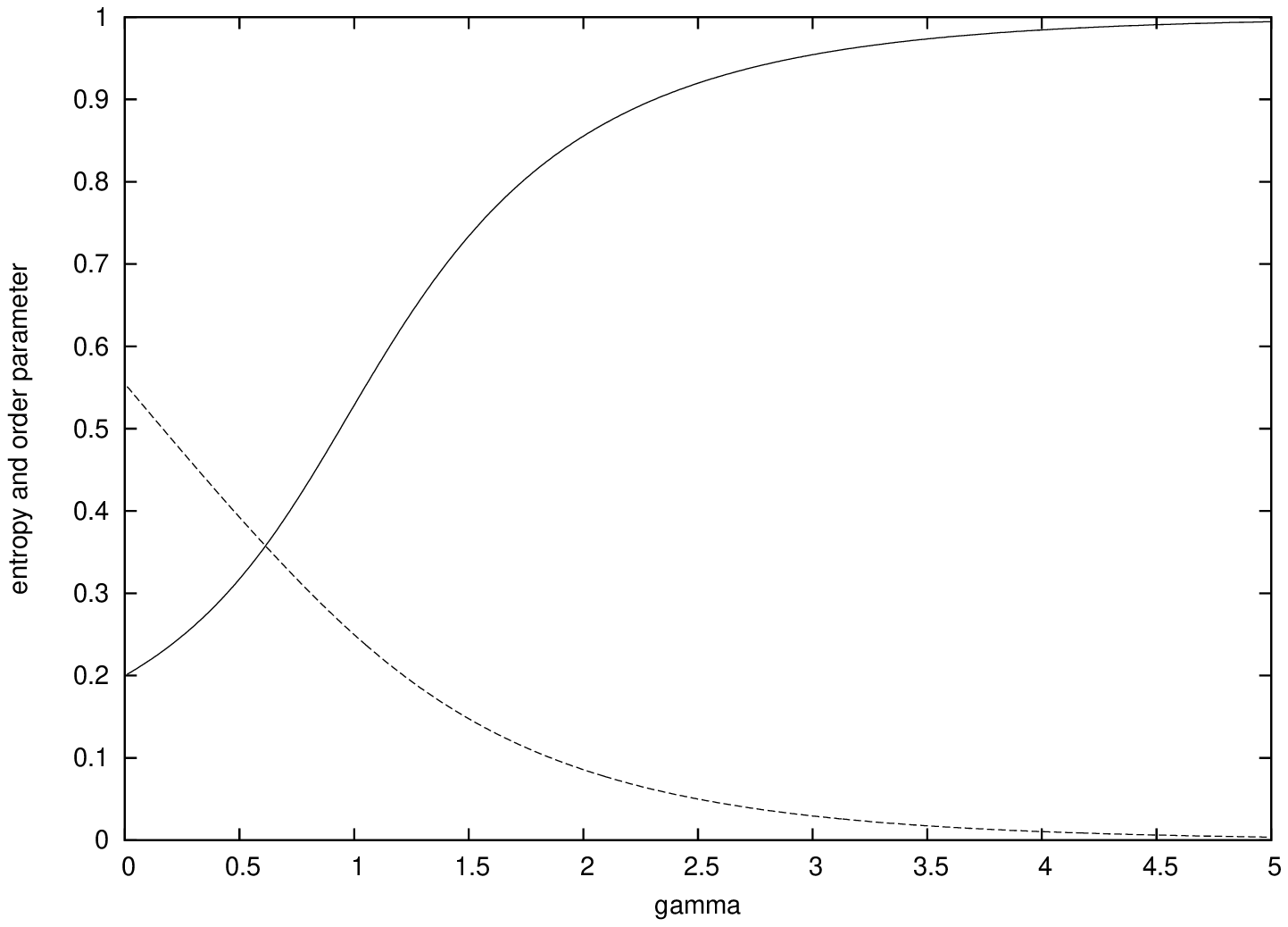,width=4.0truein,angle=270}}
\bigskip
\caption{
Entropy (dashed line) and order parameter $\theta_c$ 
(full line) as a function of
the dilution coefficient $\gamma$ at 0 temperature and 
magnetic field concentration $\delta=0.2$.
The transition disappears and the derivatives of
both entropy and $\theta_c$ are continuous everywhere.}
\label{fig2.ps}
\end{figure}

The solution can be again  found assuming that $2^n-1$ 
quantities $x(\sigma)$ take the value $\frac{1-\theta}{2^n}$
and the state  with  $\sigma^\alpha=1$ for all $\alpha$
takes the value $\frac{1+(2^n-1)\theta}{2^n}$ as in formula 
(\ref{xxpott}).

We can compute as usual and expand to the first order in $n$, than 
$\Psi_n$ is the maximum over $\theta$ of

\begin{equation}
n\log(2)\frac{\gamma}{2} 
\, \theta^2 +
n\log(2)(1-\theta)\log(1-\delta)
+ n\log(2)\delta +
n\log(2)(1-\theta) (1-\log(1-\theta))
\label{fgt}
\end{equation}

If we take into account (\ref{fpsim})
and the  large $\beta$ equalities 
$\log(\cosh(\beta)) = \beta -\log(2) \,$ and
$ \, \log(\cosh(\beta h)) = \beta h -\log(2) $, 
we can write the energy $E$ as

\begin{equation}
E = -\frac{\gamma}{2} -\delta h
\label{fem}
\end{equation}
while the entropy $S$ is

\begin{figure}
\vspace{.2in}
\centerline{\psfig{figure=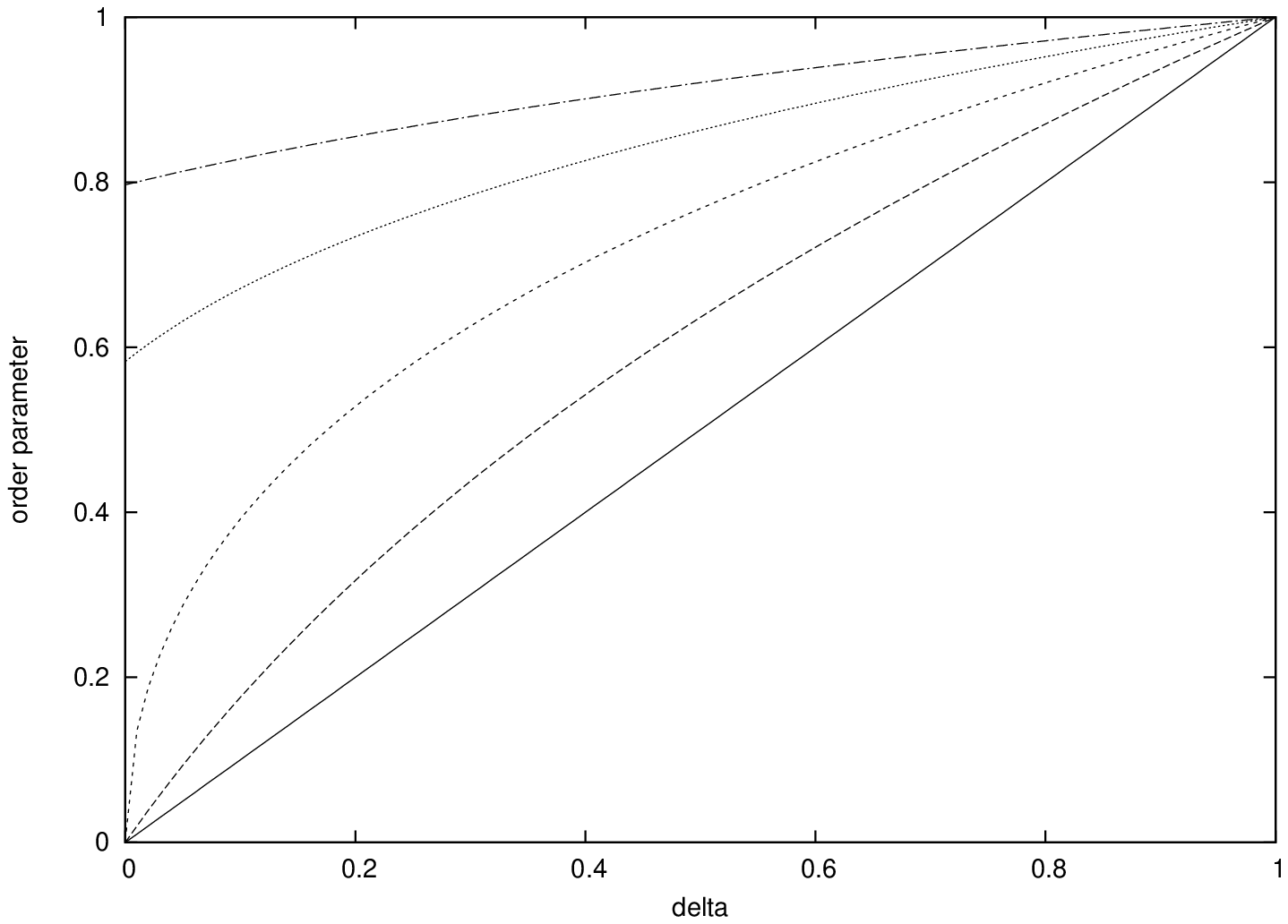,width=4.0truein,angle=270}}
\bigskip
\caption{
Order parameter $\theta_c$ as a function of 
the magnetic field concentration 
$\delta$. The different curves corresponds, starting from below,
to dilution coefficient $\gamma=0$,  
$\gamma=0.5$, $\gamma=1$, $\gamma=1.5$, $\gamma=2$.
At $\delta=0$, the order parameter $\theta_c$ vanishes only
for $\gamma \leq 1$.}
\label{fig3.ps}
\end{figure}

\begin{equation}
S= 
\log(2)\frac{\gamma}{2} 
( \theta_c^2-1)+
\log(2)(1-\theta_c)\log(1-\delta) +
\log(2)(1-\theta_c) (1-\log(1-\theta_c))
\label{feg}
\end{equation}
where $\theta_c$ is given by the equation

\begin{equation}
(1-\delta) \exp(-\gamma\theta_c) = 1-\theta_c
\label{ftheta}
\end{equation}

At variance with the 0 magnetic field case,
this equation has a single solution $\theta_c$.
The optimum parameter
$\theta_c$ is always positive and the transition disappears since
the derivatives with respect to $\gamma$ of both entropy and $\theta$ 
are continuous everywhere.
In Fig. 2 we  plot entropy and order parameter $\theta_c$ as a function of
the dilution coefficient $\gamma$ at magnetic field concentration 
$\delta=0.2$. 

In Fig. 3 we plot the order parameter $\theta$ as a function of the 
magnetic field concentration $\delta$ for 5 different values of $\gamma$:
$\gamma=0$,  $\gamma=0.5$, $\gamma=1$, $\gamma=1.5$, $\gamma=2$.
At $\delta=0$, the order parameter $\theta$ vanishes only
for $\gamma \leq 1$ where spontaneous symmetry is broken.

Finally, in Fig. 4 we plot the entropy as a function of the 
magnetic field concentration $\delta$ for the same 5 different values
of $\gamma$. The entropy is always smaller then $\log(2)$ except when both
$\gamma$ and $\delta$  vanish. Furthermore, the entropy always vanish when 
$\delta=1$ since all spins are oriented along the magnetic field.

\section{7 - Discussion}

In this paper, following Monasson~\cite{M}, 
we used magnetization densities as
replica order parameters.
The method is tested against the dilute ferromagnet model
even in the case in which a randomly located magnetic field
is present. We are able to compute exactly the
ground state energy and entropy. 

When temperature is not vanishing, we only write down the formal solution,
while the effective one asks for a correct parametrization of the 
magnetization densities $x(\sigma)$. In this case in fact, the problem
is not mapped into a simple Potts model. 
Some work in this direction is in progress.

The method can be straightforwardly applied to the dilute spin glass,
where we hope to find analogous results. 
Since, in principle, it can be used for all models,
without proliferation of order parameters,
we propose it as a general tool in the replica trick context.

\section{Acknowledgments}
We are deeply grateful to Michele Pasquini and Filippo Petroni
for their advices and suggestions which permitted many improvements.

\begin{figure}
\vspace{.2in}
\centerline{\psfig{figure=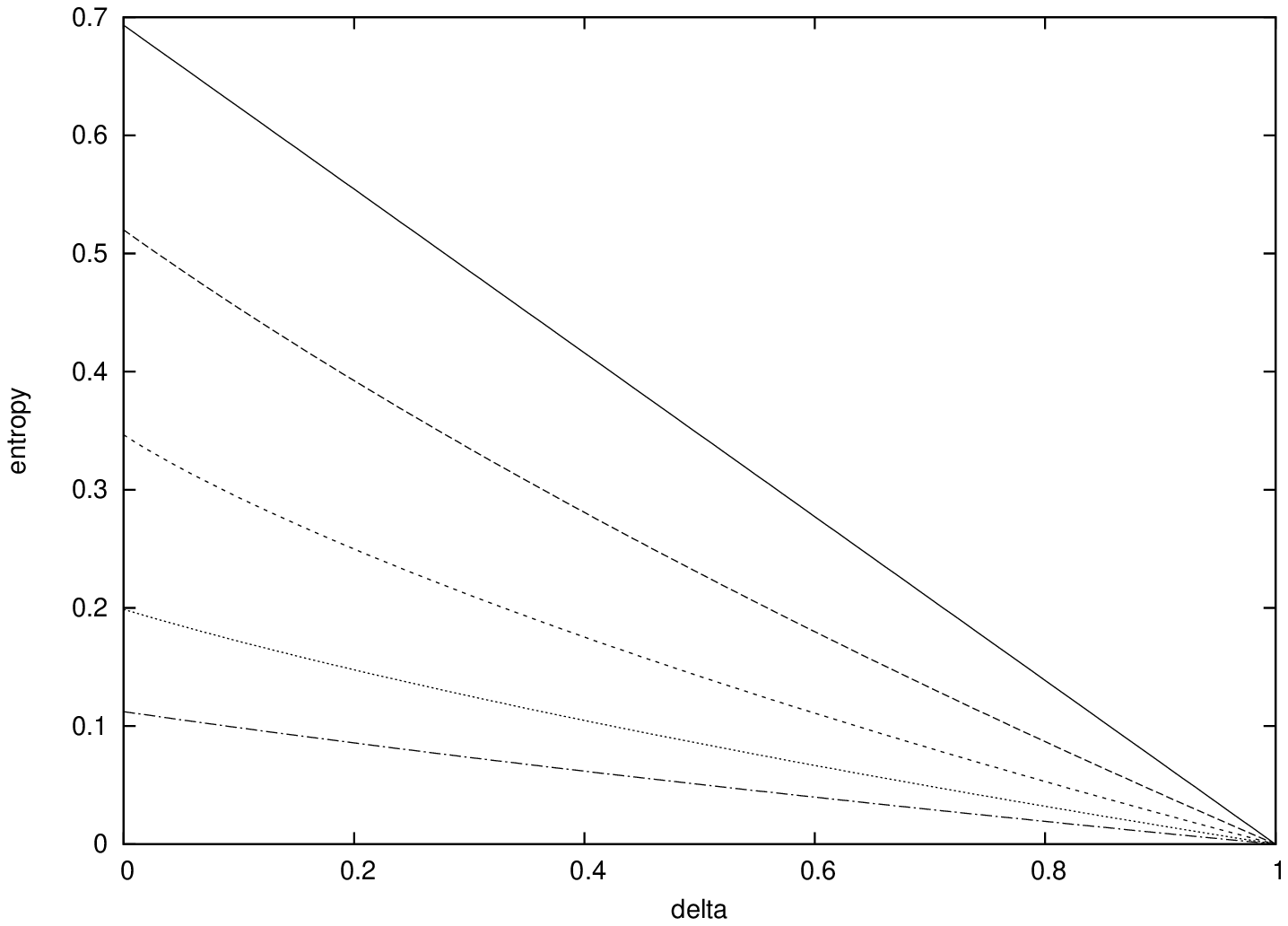,width=4.0truein,angle=270}}
\bigskip
\caption{
Entropy as a function of the magnetic field concentration 
$\delta$. The different curves corresponds, starting from above,
to dilution coefficient $\gamma=0$,  $\gamma=0.5$, $\gamma=1$, 
$\gamma=1.5$, $\gamma=2$.
At $\delta=1$ entropy is always vanishing since all spins are oriented
along the magnetic field.}
\label{fig4.ps}
\end{figure}


\begin{thebibliography}{99}

\bibitem{M}
R. Monasson, 
{\it Optimization problems and replica symmetry breaking
in finite connectivity spin glasses }, 
J. Phys A: Math. Gen. {\bf 31}, 513 (1998).

\bibitem{SK}
D. Sherrington and S. Kirkpatrick, 
{\it Solvable model of a spin-glass}, 
Phys. Rev. Lett. {\bf 35}, 1792 (1975).

\bibitem{MPV}
M. M\'ezard, G. Parisi and M. A. Virasoro,
{\it Spin glass theory and beyond}, 
World Scientific, Singapore (1987).

\bibitem{FL}
S. Franz and M. Leone, 
{\it Replica bounds for optimization problems and
diluted spin systems}, 
J. Stat. Phys. {\bf 111}, 535 (2003).

\bibitem{GT} 
F. Guerra and S. Toninelli, 
{\it 	The high temperature region of the Viana–Bray 
diluted spin glass model}, 
J. Stat. Phys. {\bf 115}, 531 (2005).

\bibitem{DSG}
L. De Sanctis and F. Guerra,
{\it Mean field dilute ferromagnet: 
High temperature and zero temperature behavior},
J. Stat. Phys. {\bf 132}, 759 (2008).

\bibitem{BC}
J. Barr\'ea, A. Ciani, D. Fanelli, F. Bagnoli and S. Ruffo,
{\it Finite size effects for the Ising model on random 
graphs with varying dilution},
Physica A {\bf 388}, 3413 (2009).

\bibitem{BG}
A. Bovier and V. Gayrard, 
{\it The thermodynamics of the Curie-Weiss model 
with random couplings}, 
J. Stat. Phys. {\bf 72} 643 (1993).

\end{thebibliography}
\end{document}